\documentclass[preprint,10pt,authoryear]{elsarticle}
\usepackage{graphicx, verbatim}

\journal{Research Policy}

\begin{document}

\begin{frontmatter}

\title{The first Italian research assessment exercise: a bibliometric perspective}

\author{Massimo Franceschet\corref{cor1}}
\ead{massimo.franceschet@dimi.uniud.it}
\ead[url]{http://users.dimi.uniud.it/~massimo.franceschet/}
\address{Department of Mathematics and Computer Science, University of Udine \\
         Via delle Scienze 206 -- 33100 Udine, Italy \\
         Phone: +39 0432 558754 / Fax: +39 0432 558499}

\author{Antonio Costantini}
\address{Department of Agriculture and Environmental Sciences, University of Udine  \\
           Via delle Scienze 208 -- 33100 Udine, Italy}

\cortext[cor1]{Corresponding author}

\begin{abstract}
In December 2003, seventeen years after the first UK research assessment exercise, Italy started up its first-ever national research evaluation, with the aim to evaluate, using the peer review method, the excellence of the national research production. The evaluation involved 20 disciplinary areas, 102 research structures, 18,500 research products and 6,661 peer reviewers (1,465 from abroad); it had a direct cost of 3.55 millions Euros and a time length spanning over 18 months. The introduction of ratings based on ex post quality of output and not on ex ante respect for parameters and compliance is an important leap forward of the national research evaluation system toward meritocracy. From the bibliometric perspective, the national assessment offered the  unprecedented opportunity to perform a large-scale comparison of peer review and bibliometric indicators for an important share of the Italian research production.  The present investigation takes full advantage of this opportunity to test whether peer review judgements and (article and journal) bibliometric indicators are independent variables and, in the negative case, to measure the sign and strength of the association. Outcomes allow us to advocate the use of bibliometric evaluation, suitably integrated with expert review, for the forthcoming national assessment exercises, with the goal of shifting from the assessment of research excellence to the evaluation of average research performance without significant increase of expenses.
\end{abstract}


\end{frontmatter}

\section{Introduction}

In December 2003 Italy started up its first-ever research assessment exercise, called \textit{Valutazione Triennale della Ricerca} (VTR), with the aim to evaluate the excellence of research activities performed by universities and other research institutions under Ministry of Education, University, and Research funding.  VTR covered the research of 20 disciplinary areas during the three-year period from 2001 to 2003. It involved the evaluation of 102 research structures including 77 universities, 12 public research agencies, and 13 private research agencies, which submitted 18,500 research products for evaluation. Peer-reviewing the submitted products involved 6,661 experts (1,465 from abroad), with a direct cost of 3.55 millions Euros and a time length of 18 months.

Evaluation activities in Italian universities traditionally favored a bureaucratic approach based on an ex ante check of the respect for input, processes, or compliance with provisions of the law \citep{MRT08}. The introduction of ratings based on the ex post quality of output and not on ex ante respect for parameters and compliance is an important cultural leap forward \citep{B98,N98}. Furthermore, the  rankings comparing the peer review ratings obtained by the universities in the different disciplinary areas were posted on the Web\footnote{\texttt{http://vtr2006.cineca.it}}. This apparently plain decision is in fact unprecedented in the setting of Italian evaluation systems in the state sector, including universities, which is characterized by a general lack of courage and a production of rankings that are based on criteria giving loose indication of merit \citep{MRT08,CG95}.

The VTR evaluation is fully based on peer review evaluation method: each submitted research product was assessed by a pool of experts who expressed a qualitative judgement that is then mapped to a quantitative categorial rating.  \citet{RBC07} show that the VTR exercise was carried out on the basis of assessment criteria proposed in the literature for peer-reviewing (rationality, reliability, impartiality, efficiency, effectiveness), controlling the presence and the relevance of bias of the peer judgements (prestige of institutions and reputation of scientists).  Hence, we assume here that peer reviewers expressed a reliable judgement on the products submitted at VTR and that this rating reflects the intrinsic \textit{quality} of the product.

Submitted products were autonomously selected by research institutions in the measure of at most one product every four researchers (universities) or every two researchers (research agencies) choosing among the entire production over a three-year period. In order to maximize peer rating, each structure selected the products deemed to be of highest quality. It turned out that, for areas in which journal publication is the routine, most of the submitted products are journal articles and most of these articles appear in journals indexed in databases of Thomson Reuters, formerly known as ISI\footnote{At the moment, Thomson Reuters Web of Science, Elsevier Scopus, as well as Google Scholar are the main multi-disciplinary bibliometric data sources.}. For each article covered by Thomson Reuters, we have at disposal an \textit{article citation rating}, measuring the number of citations that the article received from other papers in the database, and a \textit{journal citation rating}, evaluating the impact factor of the journal in which the article appears, which corresponds to the average number of recent citations received by papers published in the journal \citep{GS63}. Furthermore, for relatively large publication sets, we may compute the recently proposed and highly celebrated Hirsch index, which attempts to assess both production and impact in a single figure \citep{H05,Ba07}.

This opens the unprecedented opportunity to perform a large-scale comparison of peer review and bibliometric indicators for the Italian research system. This is the aim of the present contribution. More specifically, we pose the following \textit{research questions}:

\begin{enumerate}

\item Are peer review judgements and (article and journal) bibliometric indicators independent variables?

\item If not, what is the strength of the association?

\item In particular, is the association between peer judgement and article citation rating significantly stronger than the association between peer judgement and journal citation rating?
\end{enumerate}

Answering these questions is of crucial importance to evaluate the opportunity of using bibliometrics in the next research assessment exercises.

In Section \ref{vtr} we concisely describe the VTR assessment exercise. In Section \ref{study} we address the posed questions with a careful analysis comparing peer review and bibliometric indicators at both levels of research disciplines (Section \ref{disciplines}) and research structures within disciplines (Section \ref{structures}). Related work is amply surveyed in Section \ref{related}. Finally, in Section \ref{conclusion} we draw some conclusions.

\section{An overview of VTR} \label{vtr}

VTR was managed by the Committee for the Evaluation of Research (CIVR) and was designed as an ex post assessment exercise based on \textit{peer review}. Its plan can be summarized as follows.
CIVR divided the national research system into 20 scientific-disciplinary areas, including 6 interdisciplinary sectors, and set up an evaluation panel responsible for the assessment of each area. Panels were composed by high level experts (panelists), which number fluctuated from 5 to 17 according to the area size and disciplinary complexity. The exercise was then articulated in three phases, that were in charge of research structures, panels and CIVR, respectively.

In the initial phase, research institutions submitted to panels a set of autonomously selected research products. Types of products admitted to submission are: journal articles, books, book chapters, proceedings of national and international conferences, patents, designs, performances, exhibitions, manufactures and art operas. The only mandatory principle of selection stated that products of research should not exceed 50\% of the full-time-equivalent researchers in the institution.\footnote{A full-time-equivalent researcher represents 0.5 researchers in universities, where researchers teach as well, while it corresponds to 1 researcher in research agencies. Hence, universities were allowed to submit a maximum number of products corresponding to 25\% of the three-year average permanent academic staff.} The research structures submitted an overall sample of 18,500 products partitioned as follows:  journal articles 72\%, books 17\%, book chapters 6\%, patents 2\% and the remaining typologies 3\%. Evaluated products were more than 17,300 (there are products submitted by more than one institution). Research structures were also demanded to transmit to CIVR data and indicators about human resources, international mobility of researchers, funding for research projects, patents, spin-off and partnerships, allowing to reveal impact on employment.

In the second phase of the exercise, which was carried out with the aid of a web platform, panelists assigned research products to external referees. Each product was assessed by at least two referees who peer-reviewed it according to four aspects of merit: quality (the opinion of peer on the scientific excellence of the product compared to the international standard), importance, originality and internationalization. Referees also expressed a final score on the following four-point scale:

\begin{enumerate}

\item \textit{excellent}: a product within the top 20\% of the value in a scale shared by the international scientific community;

\item \textit{good}: a product in the 60\%-80\% segment;

\item \textit{acceptable}: a product in the 40\%-60\% segment;

\item \textit{limited}: a product within the bottom 40\%.

\end{enumerate}

For every evaluated product panels drew up a consensus report where panelists re-examined the peer judgments and fixed the final score. Furthermore, CIVR weighted the peer review scores as follows: 1 (excellent), 0.8 (good), 0.6 (acceptable), and 0.2 (limited). The numeric formulation made it possible to sum product scores, in order to obtain a mean rating for single research structures  providing a proxy for the value of the institution research performance and the possibility to compile corresponding rankings of structures. Rankings were compiled for each disciplinary area and within groups of structures of comparable sizes: mega structures (more than 74 products), large structures (25-74 products), medium structures (10-24 products), and small structures (less than 10 products). Panels provided a final report including ranking lists of the institutions in the surveyed area, highlighting strength and weakness points of the research area, and proposing possible actions of improvement.

In the final phase of the assessment exercise, CIVR produced a detailed analysis of requested data and indicators, integrating panel reports with collected data about human resources and project funding. The CIVR final report defines a first-ever comprehensive assessment of the national research system. In summer 2009, VTR outcomes have been used for the first time by Ministry of Education, University, and Research to allocate a 7\% share of the Ordinary Fund for Higher Education (FFO).

\section{A bibliometric analysis of VTR} \label{study}

Our analysis considers the following research areas:

\begin{enumerate}

\item mathematics and computer sciences (MCS);
\item physics (PHY);
\item chemistry (CHE);
\item earth sciences (EAS);
\item biology (BIO);
\item medical sciences (MED);
\item agricultural sciences and veterinary medicine (AVM);
\item civil engineering and architecture (CEA);
\item industrial and information engineering (IIE);
\item economics and statistics (ECS).

\end{enumerate}

We excluded from our investigation the six interdisciplinary areas as well as the following four areas: philological-literary sciences, antiquities and arts; history, philosophy, psychology and pedagogy; law; political and social sciences. The number of submitted products in these areas that are covered by Thomson Reuters databases is too modest for a reliable application of bibliometrics.

In the following, we refer to a product contained in Thomson Reuters databases as a Thomson Reuters (TR) article. For each submitted product we have at disposal a peer review judgement. Moreover, for each TR article we computed the following bibliometric indicators:

\begin{enumerate}
\item \textit{article citation rating}, counting the number of citations that the article received from other TR papers. We retrieved all citations recorded in Thomson Reuters Web of Science database received by more than 17,000 papers up to June 2006. Since papers refer to period 2001-2003, this means that we used a citation window of minimum length of 2.5 years, maximum length of 5.5 years, and average length of 4 years. These periods are generally sufficient for a paper to collect the peak of citations in each of the surveyed disciplines;

\item \textit{journal citation rating}, evaluating the average number of recent citations received by papers published in the journal in which the article appears. We computed the average 2-year journal impact factor over the period 2001-2003.

\end{enumerate}

Furthermore,  we computed the Hirsch (h) index over relatively large sets of papers. The h index for a publication set is the highest number $n$ such that there are $n$ papers in the set each of them received at least $n$ citations \citep{H05}. The h index immediately found interest in the public \citep{Ba07} and in the bibliometrics literature (see \citet{BD07b} for opportunities and limitations of the h index). In particular, it is currently computed by both Thomson Reuters Web of Science and Elsevier Scopus bibliometric data sources. The index is meant to capture both production and impact of a publication set in a single figure. It favors publication sets containing a continuous stream of influential works over those including many quickly forgotten ones or a few blockbusters. Moreover, the index is robust to self-citations: all self-citations to papers with less than h citations are irrelevant for the computation of the index, as are the self-citations to papers with many more than h citations.

We aggregated peer review and bibliometric data at both levels of research disciplines (Section \ref{disciplines}) and research structures within disciplines (Section \ref{structures}).

\subsection{Analysis at the level of research discipline} \label{discipline}

\begin{table}[t]
\begin{center}
\begin{tabular*}{1\textwidth}{@{\extracolsep{\fill}}lrrrrrrrr}
\textbf{area} &	\textbf{size} & \textbf{cov} & \textbf{auth} & \textbf{own} & \textbf{peer} &	 \textbf{cites}  & \textbf{IF} &  \textbf{h}\\ \hline
\textbf{MCS} &  787   & 92\% & 2.26 & 69\% & 0.830 (0.831) & 3.97  (3.54) & 1.12 &   18 \\ \hline
\textbf{PHY} &  1767  & 89\% & 51.85& 42\% & 0.879 (0.885) & 24.66 (4.26) & 5.79 &   87 \\ \hline
\textbf{CHE} &  1089  & 92\% & 5.10 & 68\% & 0.807 (0.813) & 16.14 (3.14) & 5.14 &   50 \\ \hline
\textbf{EAS} &   651  & 90\% & 4.17 & 64\% & 0.825 (0.836) & 7.33  (2.44) & 3.01 &   26 \\ \hline
\textbf{BIO} &  1575  & 96\% & 6.56 & 66\% & 0.826 (0.831) & 24.58 (2.90) & 8.48 &   83 \\ \hline
\textbf{MED} &  2639  & 96\% & 8.47 & 59\% & 0.776 (0.780) & 26.65 (3.20) & 8.34 &   106 \\ \hline
\textbf{AVM} &   750  & 89\% & 4.81 & 67\% & 0.712 (0.728) & 8.20  (3.08) & 2.66 &   27 \\ \hline
\textbf{CEA} &   758  & 45\% & 2.40 & 84\% & 0.750 (0.755) & 3.58  (3.10) & 1.16 &   14 \\ \hline
\textbf{IIE} &  1195  & 82\% & 3.48 & 77\% & 0.774 (0.779) & 4.78  (2.98) & 1.61 &   23 \\ \hline
\textbf{ECS} &   971  & 54\% & 1.86 & 76\% & 0.673 (0.799) & 3.16  (3.63) & 0.87 &   17 \\ \hline
\end{tabular*}
\end{center}
\caption{Analysis at the level of research discipline.}
\label{disciplines}
\end{table}

Table \ref{disciplines} contains, for each surveyed discipline, the following columns:

\begin{itemize}
\item \textit{area}: the disciplinary area abbreviated as above;

\item \textit{size}: the number of submitted products.\footnote{Papers with authors affiliated to structures belonging to different areas are counted for each affiliation area.} This gives an indication of the size (number of researchers) of the field;

\item \textit{cov}: the fraction of submitted products that are covered in TR databases;

\item \textit{auth}: the mean number of authors per paper. We interpret this as a measure of discipline propensity of collaboration among scholars;

\item \textit{own}: degree of ownership. For a given paper submitted by a given structure, it is the number of paper authors that are affiliated to the structure that submitted the paper divided by the total number of paper authors. It demonstrates the discipline propensity of collaboration with scholars of different research structures (belonging to the same or different fields): the lower the degree of ownership, the higher the inter-structure collaboration propensity.

\item \textit{peer}: the average peer review rating. Within brackets we show the rating over TR articles only;

\item \textit{cites}: the average number of received citations. Within brackets we show the ratio between number of citations and impact factor;

\item \textit{IF}: the average impact factor of the journals publishing the papers;

\item \textit{h}: the h index for the set of submitted TR papers.

\end{itemize}

The largest area is MED, followed by PHY and BIO; small fields are EAS, AVM, CEA and MCS.
All areas have a large TR coverage with two notable exceptions: CEA (45\%) and ECS (54\%); important sub-fields of these areas frequently publish on books, which are not covered by TR. More precisely, CEA groups civil engineering and architecture; the former mostly publish in journals and has a good TR coverage (75\%). On the contrary, scholars in architecture frequently publish books and book chapters and hence the TR coverage is limited (3\%). It follows that, for the purpose of this study, the output of area CEA is largely dominated by civil engineering products. As for ECS, it is mainly composed of economics, management, and mathematics. Scholars in mathematics and economics publish mostly in journals, but these are differently covered by TR (56\% in economics versus 78\% in mathematics). Scholars in management prefer books or book chapters, reducing the TR coverage to 22\%. Computer scientists typically prefer conference proceedings to archival journals as a mean of publication but typically journals convey a higher impact \citep{F10-SCIENTO,F10-CACM}. Although TR does not index conference proceedings (at least it did not at the time of the assessment exercise), TR coverage of MCS is reasonably high. This because computing structures submitted for evaluation mostly journal papers instead of the more frequent proceeding papers, probably because they perceived that these publications are of higher quality.

The mean number of authors varies across disciplines. PHY, MED, and BIO are the fields with the largest number of authors per paper, while ECS, MCS, CEA are the areas with the lowest authorship propensity. Notice that PHY is a significative outlier: on average, papers in this discipline have more than 50 authors. A closer look to the authorship distribution reveals that it is highly skewed: there are many papers with few authors and few ones with a huge number of authors. The median number of authors is 5, meaning that at least 50\% of the papers have at most 5 authors, a figure comparable with other disciplines. On the other hand, 13\% of the papers have more than 100 authors, and there exists a hub paper with the impressive number of 1412 co-authors. This phenomenon, known as \textit{hyperauthorship} and typical of certain areas of research including high energy physics, is investigated in \citet{C01}.

We observed a significative negative correlation between authorship and ownership\footnote{Spearman coefficient -0.82, p-value 0.007.}: the larger the number of authors per paper, the lower the ownership degree of papers, indicating a stronger propensity to collaborate outside the home institution. For instance, more than half of the authors of papers in PHY belong to a different structure with respect to the submitting one. At the other extreme, authors in CEA and, to a less extent, those in IIE and ECS, prefer to work in small groups within their research structures.

Peer review judgements were, on average, quite high, reflecting the selection of the best papers only provided by each structure in each discipline. Moreover, the average judgement over all products corresponds to the mean judgement with respect to TR articles only, with the exception of area ECS: in this field TR articles have been evaluated significantly higher than non-TR products. The fields with the best peer ratings are PHY, MCS, BIO, and EAS in this order. The areas with the poorest peer judgements are ECS and AVM. In the case of economics and statistics, an explanation of the bad performance is the high frequency of non-TR products which received a low peer rating. Furthermore, \citet{RBC07} claim that the lower levels of rating for this area are also associated with the higher disagreement of the panel consensus in this sector with respect to the others. As for AVM, the ratings of its sub-fields are: agronomy (0.678), entomology (0.681),
veterinary science (0.684), food and nutrition (0.697), animal science (0.720), plant science
(0.721), agricultural chemistry (0.757). Based on available ratings, animal science, plant science e agricultural chemistry tend to be in line with situations of good scientific quality, but the other sub-fields rank below the national standard.

Bibliometric indicator scores wildly vary across fields. This field effect is a well known phenomenon in bibliometrics (see, e.g., \citet{AWBB08}). This is mainly due to the different field publication coverage of the underlying bibliographic databases and to the different field citation habits, including number of references per paper and citation speed. The ratio between article citation and impact factor scores is supposed to mitigate the field effect. It tells us something about the ability of the institutions from different fields to select the papers with the highest potential impact. In this respect, PHY was the best area and EAS was the worst.

Tables \ref{correlation1} and \ref{correlation2} in Appendix analyze the variables size (number of papers), average number of citations per paper, average journal impact, and h index across sets of papers characterized by different levels of (peer review) quality. The size factor gives more insight into the area overall peer judgement (column peer in Table \ref{disciplines}). For instance, papers in PHY received the highest peer review judgements (0.879 on average). Indeed, more than half (52\%) of them have been judged  excellent, while only 1\% of them  have been considered  limited products. On the other hand, peer reviewers were very critical with respect to products in ECS (the average rating is 0.673): only 17\% of the products in this area are excellent works, and a higher share, 18\%, are considered limited contributions. Notice that, for all areas but PHY, the most popular referee opinion is good.

Article citations are positively correlated with the categorial peer review judgement: generally, the average number of citations per paper decreases as the peer rating declines. Excellent papers always receive the highest average number of citations, well above the discipline mean, while acceptable and limited contributions received an average citation impact lower than the discipline mean. Nevertheless, some exceptions to positive correlation exist, namely the impact of limited products in EAS (+2 positions in the categorial ranking), MED (+1), CEA (+1), and IIE (+2).

Journal impact factors are also positively correlated with peer assessment: on average, the impact factor of publishing journals drops as the peer evaluation decreases. The association is, however, not as strong as the one noticed for article citations. Indeed, there are more exceptions to  positive association, namely the acceptable papers in MCS (+1 positions in the categorial ranking) and EAS (+1), and the limited products in PHY (+2), MED (+2), CEA (+1), and IIE (+2).

The h index discriminates very well between different peer review ratings with only two exceptions: excellent and good papers in fields AVM and CEA. In particular, the h index neatly separates the lower judgements acceptable and limited, on which the discrimination power of both article and journal citation measures is weaker. Take, for example, the sets of acceptable and limited papers in MED. Both the average number of paper citations and the average journal impact factor for limited articles are above the same measures for acceptable papers. On the other hand, the h index of acceptable papers (36) largely dominates that of limited articles (21). Indeed, the sorted citation sequence for acceptable publications features a longer stream of influential papers while that for limited papers is headed by two blockbusters, which are responsible for the relatively high mean citation values, but then it quickly decreases.

The relationship between peer judgements and bibliometric indicators, in particular article and journal citation indices, has been further investigated. Within each discipline, we expressed the discrete variable article citation as a categorial variable by splitting the distribution into quartiles to obtain a four-point scale for the variable. We did the same for journal impact factor. Then, we prepared, for each discipline, a contingency table displaying the categorial variables peer judgement and article citation (Tables \ref{crossCites1} and \ref{crossCites2} in Appendix) and a similar table for peer judgement and journal impact factor (Tables \ref{crossIF1} and \ref{crossIF2} in Appendix). Each table cell contains the joint relative frequency for the conditional distribution of the bibliometric variable (either article citation or journal impact factor)  given the peer judgement variable. For example, Table \ref{crossCites1}, discipline BIO, shows that excellent papers in the discipline are split into citation quartiles as follows: 11.3\% in the 1st quartile, 18.4\% in the 2nd quartile, 25.3\% in the 3rd quartile, and 45.0\% in the 4th quartile.

It turns out that, with very few exceptions,  the majority of excellent papers are associated with the highest bibliometric quartile (the 4th one), while the majority of limited products belong to the lowest bibliometric quartile (the 1st one). Good and acceptable papers distribute over the four quartiles, with a preference for the lower segments, in particular for acceptable products.
If bibliometric and peer assessments were independent variables, we would expect that the relative frequency of each cell would be the product of its marginals (the row and column relative frequencies). Hence, we can test the independence of the bibliometric and peer review variables by comparing the observed frequencies with the expected ones in case of independent variables (this is the well-known  Pearson chi-square test for independence). The output of the test is that, for all disciplines, peer judgement and bibliometric indicators are \textit{not} independent variables (with a significance level less than 0.001) with the unique exception of journal impact factor for area MCS. The strength of the association between peer opinion and article citation variables, measured with Spearman's rank-order coefficient, ranges from 0.187 for IID to 0.403 for PHY. All values are significantly different from 0 (p-value $<$ 0.001). The association between peer judgement and journal impact factor ranges from 0.197 for IID to 0.529 for AVM. All values except that for MCS are significantly different from 0 (p-value $<$ 0.001).

To conclude the investigation of association between peer review and bibliometrics, we performed a probabilistic analysis (Tables \ref{probability1} and \ref{probability2} in  Appendix). Namely, for each pair of adjacent peer judgments $X$ and $Y$, we computed the  probability $P(c(X) > c(Y))$ (respectively, $P(c(X) = c(Y))$) that for two randomly drawn papers $P$ and $Q$ rated $X$ and $Y$, respectively, the number of citations of $P$ is greater than (respectively, equal to) the number of citations of $Q$. If peer judgments are positively correlated with article citations, an educated guess would be that, if rating $X$ is above $Y$, then $P(c(X) > c(Y))$ is larger then $P(c(Y) > c(X))$. It holds that $P(c(X) > c(Y))$ can be expressed as the following ratio:
$$
P(c(X) > c(Y)) = \frac{|\{(P,Q).\ r(P) = X \ \mathrm{and} \ r(Q) = Y \ \mathrm{and} \ c(P) > c(Q)\}|}{|\{(P,Q).\ r(P) = X \ \mathrm{and} \ r(Q) = Y\}|}
$$
where $r(P)$ is the rating of $P$, $c(P)$ is the number of citations received by $P$, and $|\cdot|$ is the cardinality of a set. Clearly, we have that $$P(c(X) > c(Y)) + P(c(Y) > c(X)) + P(c(X) = c(Y)) = 1$$ Similarly we computed the probabilities $P(IF(X) > IF(Y))$ and $P(IF(X) = IF(Y))$ for the journal impact factor.

\begin{figure}[t]
\begin{center}
\includegraphics[scale=0.35, angle=-90]{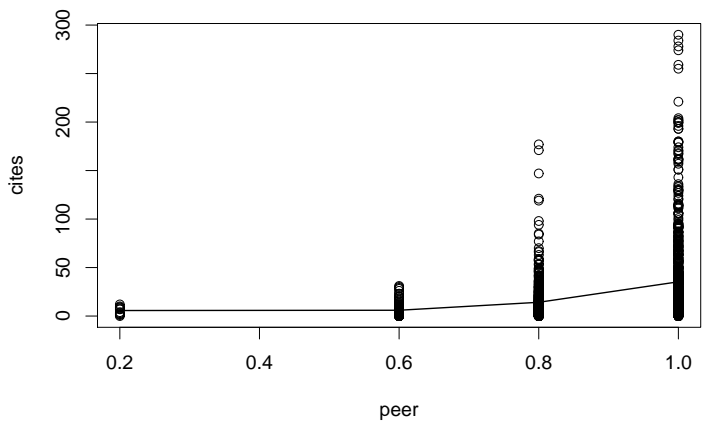}
\caption{Categorial scatter plot showing citations received by papers of different peer-assigned quality for research area BIO. The solid line connects the mean number of citations for each group. Papers of higher quality generally receive more citations.}
\label{categorial}
\end{center}
\end{figure}

We observe that for pairs of judgements (E,G) and (G,A), the number of pairs of articles whose citations are concordant with the judgements is always greater than the number of discordant pairs of papers: the higher peer rating, the higher the probability of finding highly cited papers as well as that of finding papers published in journals of high impact.  For the rating pair (A,L) the situation is more controversial: in four cases over ten, the exploited bibliometric indicators are less accurate at distinguishing acceptable papers from limited ones. By way of example, Figure \ref{categorial} illustrates the found association between citations and peer assessment for research area BIO. The probability that an excellent paper receives more citations than a good one is 0.68 (as opposed to 0.30 for the probability of the opposite event), the probability that a good paper collects more citations than an acceptable one is 0.64 (as opposed to 0.33), and the probability that an acceptable paper harvests more citations than a limited one is 0.59 (as opposed to 0.35). Furthermore, in 88\% of the cases a paper rated excellent receives more citations than a paper judged limited, while in only 10\% of the cases the opposite happens (2\% of the times the two papers receive the same number of citations).

\subsection{Analysis at the level of research structures} \label{structures}

In this section we investigate the structure rankings within each disciplines compiled with respect to peer review judgements and bibliometric indicators. For the sake of statistical significance, for each discipline, we included in this analysis only research entities that submitted at least 10 products belonging to the discipline. For each structure in each discipline we compute the following ratings:

\begin{itemize}
\item \textit{peer review rating}: this is the average peer review judgment of the products submitted by the structure; we also consider the peer review judgment restricted to TR articles;
\item \textit{article citation rating}: this is the average number of citations received by TR articles submitted by the structure;
\item \textit{journal citation rating}: this is the average impact factor of journals that published the TR articles submitted by the structure.
\end{itemize}

Universities were allowed to submit a maximum number of products corresponding to (only) 25\% of the three-year average permanent academic staff. Research institutions were partitioned according to the number of submitted products in mega structures (over 74 submitted products), large structures (from 25 to 74 products), medium structures (from 10 to 24 products), and small structures (less than 10 products). Except for mega structures, the numbers of submitted products are, in general, not sufficient for a reliable computation, at the structure level, of the h index, whose score, by definition, is bounded by the number of papers in the evaluation set. For this reason, we do not consider the h index in the present analysis at the level of research structures.

We performed a rank-order correlation analysis to compare the structure compilations according to peer review and bibliometric ratings.  We tested the hypothesis that the Spearman correlation coefficient is different from null and, when it holds, we investigated the strength of the correlation. Table \ref{spearman} gives the main outcomes for the analysis. The used peer rating refers to TR articles only. The outcomes are summarized in the following:

\begin{table}[t]
\begin{center}
\begin{tabular*}{1\textwidth}{@{\extracolsep{\fill}}l|lr|lr}
\hline
\textbf{area} &	\textbf{peer vs.\ cites} && \textbf{peer vs.\ IF} \\
&	$\sigma$ & p-value  & $\sigma$ & p-value \\ \hline
\textbf{MCS} &  0.46 & 0.015    & 0.52 & 0.005     \\ \hline
\textbf{PHY} &  0.81  & $<$0.001 & 0.29  & 0.088   \\ \hline
\textbf{CHE} &  0.60  & $<$0.001 & 0.85  & $<$0.001 \\ \hline
\textbf{EAS} &  0.79  & $<$0.001 & 0.34  & 0.140    \\ \hline
\textbf{BIO} &  0.69  & $<$0.001 & 0.74  & $<$0.001  \\ \hline
\textbf{MED} &  0.56  & $<$0.001 & 0.60  & $<$0.001  \\ \hline
\textbf{AVM} &  0.52  & 0.015    & 0.52  & 0.015     \\ \hline
\textbf{CEA} &  0.32  & 0.124    & 0.41  & 0.043     \\ \hline
\textbf{IIE} &  0.58  & $<$0.001 & 0.38   & 0.036    \\ \hline
\textbf{ECS} &  0.42  & 0.006    & 0.45   & 0.003   \\ \hline
\end{tabular*}
\end{center}
\caption{Rank-order correlation between structure rating variables: peer review rating of TR articles (peer) is compared to article citation rating (cites) and to journal citation rating (IF). We show the Spearman rank-order correlation coefficient ($\sigma$) and the significance of the test (p-value).}
\label{spearman}
\end{table}

\begin{itemize}
\item There is an overall positive correlation between peer rating and  article citation rating at the structure level. In particular, for six areas, namely PHY, CHE, EAS, BIO, MED, and IIE, the correlation is significant at a level less than 0.001, and for MCS, AVM, and ECS the correlation is significant at a level of 0.02.\footnote{A correlation is considered significant when the p-value is less than or equal to 0.05.} On the other hand, the correlation is not significant for area CEA.

\item A highly significative correlation between peer rating and journal citation rating is less frequent: in only cases, CHE, BIO, and MED, the correlation is significative at a level less than 0.001. The association is significant at the level of 0.05 for areas MCS, AVM, CEA, IIE, ECS. The correlation is not significant for areas PHY and EAS, which are the areas with the highest association with respect to article citation.
\end{itemize}

\begin{figure}[t]
\begin{center}
\includegraphics[scale=0.35, angle=-90]{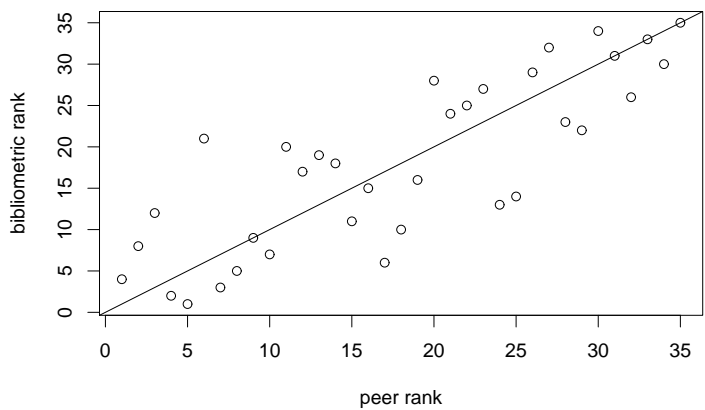}
\caption{Rank plot comparing peer and article citation ratings for structures in research area BIO. For each structure, the rank of the structure according to peer rating is plotted against the structure rank according to article citation rating. Peer rating favors structures above the solid bisector line and hampers those below, while those on the line do not change their ranks in the two compilations.}
\label{rankplot}
\end{center}
\end{figure}

By way of example, Figure \ref{rankplot} contains a rank plot comparing peer and article citation ratings for structures in research area BIO (35 structures that submitted at least 10 products). In general, the structure rank in the citation compilation increases as the structure rank in the peer compilation rises. The median change of rank is 4 (11\% of the compilation length). Peer review, compared to citation rating, mostly favours structures Milano (+15 positions with respect to the citation compilation), Trento (+9), L'Aquila (+9), and Roma Tre (+8). On the other hand, structures that are most advantaged by using the bibliometric ranking are Genova (+11 positions with respect to the peer compilation), Pavia (+11), and Trento (+11). Institutions Roma La Sapienza, Palermo, ENEA, and Parma do not change their positions in the two listings.

\section{Related work} \label{related}

The most famous and discussed European national research evaluation is the Research Assessment Exercise (RAE) in Great Britain, which started in 1986. It is a peer review evaluation dealing with approximately half of the total portfolio of research outputs of the assessed institutions -- for RAE 2008, research structures were invited to submit four research products for each full-time researcher (as opposed to one product every four researchers in the Italian RAE). Interestingly, it was announced that, after 2008 edition of the exercise, a system of citation-based metrics will be introduced to inform and supplement peer review where robust data are available -- most likely in medicine, science, and engineering -- with the goals of achieving consistency, international benchmarking, and where possible reducing workloads. The first new exercise, renamed Research Excellence Framework (REF), is due to be completed in 2013.

In the US, evidence suggests that publication and citation metrics are more readily accepted and more liberally applied. \citet{C96} cites the following example to illustrate the greater tolerance of evaluative bibliometrics in North America:

\begin{quote}
\textit{In a recent legal action initiated by a female assistant professor of biology, who had been denied tenure at Vassar College, the plaintiffs lawyer brought forward as evidence of discrimination the fact that her untenured client had a higher citation count than some tenured male staff in the same department. Although the female candidate's case was overturned subsequently on appeal (in part, and ironically, as a result of errors in the citation data submitted as evidence), the legal admissability and potential courtroom impact of citations are worthy of note.}
\end{quote}

The literature offers more than a few contributions dedicated to the comparison of peer review and bibliometric evaluation methodologies. The following is a (necessarily incomplete) selection. The use of citation metrics in place of, or as a supplement to, the UK RAE has been considered extensively. For instance, \citet{ON03} observe a statistically significant correlation between the 2001 RAE result and citation counts for archeology, and contains references to other studies that have found positive associations for other fields and exercises.

\citet{vR06c} investigates the statistical correlation between different bibliometric indicators, including the h index and the `crown indicator' (a citation average normalized to world average, a measure developed and implemented by the author's group at Leiden) with peer review judgement for university chemistry research groups in the Netherlands. Results show that the h index and the crown indicator both relate in a quite comparable way with peer judgements. In particular, both indicators discriminate very well between highly rated groups and poorly rated ones, but less well between good and excellent judgements.

\citet{BD07} investigate the convergent validity of decisions for awarding long-term fellowships to
post-doctoral researchers as practiced by the Boehringer Ingelheim Fonds -- an international
foundation for the promotion of basic research in biomedicine -- by using the h index. Grant and fellowship peer review is principally an evaluation of the potential of the proposed research. The h indices of approved applicants are on average consistently higher than those of rejected applicants. Nevertheless, the distributions of the h indices partly overlap: some rejected applicants have a h index that is substantially higher than that of approved applicants, and some approved applicants have a h index that is substantially lower than that of rejected applicants.

\citet{RLVR98} study the correlation between bibliometric indicators and the outcomes of
peer judgements of research programmes made by expert committees of condensed matter physics in the Netherlands. In particular, a breakdown of correlations to the level of different peer review criteria has been made. The authors draw a number of interesting conclusions. Positive and significant but no perfect correlations are found between a number of bibliometric indicators (in particular average number of citations per publication and the above mentioned crown indicator) and peer judgements of research programmes. The impact of publication journals, as reflected by the mean journal citation rates, does not correlate well with the quality of these programmes as perceived by peers. A negative correlation is found between the percentage of self-citations and jury ratings. Correlations between bibliometric indicators and expert judgements are higher in the case of `curiosity driven' basic research than in the case of `application driven' research. Finally, at the level of specific criteria used by juries, the highest correlation is found between
ratings for bibliometric indicators and the criterion `team' -- the assessment of the competency of researchers and of the research team.

\citet{AT04} investigate the relationship between bibliometric indicators and the outcomes of peer reviews based on a case study of research groups within the natural sciences at the University of Bergen, Norway. The analysis shows positive but relatively weak correlations. Groups obtaining the highest citedness indices were all rated as very good or excellent. On the other hand, groups cited below the world average obtained rather heterogeneous ratings. The authors conclude that peer review and bibliometric methods should be used in combination. In particular, in cases where there is a significant deviation between the two evaluation outcomes, the panel should investigate the reasons for these discrepancies.

The preceding comparisons are limited to only a few disciplinary sectors or to just one sector, or even to a single institution. By contrast, our investigation spans over 10 disciplinary areas in the sciences and social sciences and involves the output of more than a hundred public and private research structures. Two contributions mostly relate to ours. \citet{RBC07} analyse the output of Italian VTR for four areas: chemistry, biology, economics and humanities. The authors find a general consensus between expert advice (but weaker in economics) and show that peer review was not biased toward prestige of institutions or reputation of scientists. On the other hand, they notice a bias linked to the interdisciplinary (non-conventional) research. Furthermore, the authors perform a Spearman correlation analysis as well as an ordinal regression one to compare peer judgements of papers with the impact factor of journals publishing the papers for chemistry, biology, and economics areas. They find a statistically significant association, although not strong, and conclude that ``\textit{this reinforces the idea that impact factor is a good predictor of the quality of journals -- not for the quality of articles published in a particular journal}". Finally, they suggest that ``\textit{further developments of VTR should go toward a larger use of the bibliometric indicators, in conjunction with peer review}".

\citet{ADC09} provide a broader investigation on Italian VTR outcomes for eight disciplines, the ten disciplines we have used in our study with the exclusion of civil engineering and architecture (CEA) and economics and statistics (ECS), for which the database coverage is less important. The authors correlate, at the research structure level, peer quality opinions on papers with metrics based on the impact factor of the journals publishing the papers, normalized across scientific disciplinary sectors within disciplinary areas. They conclude that the two evaluation methods (peer review and bibliometrics) significantly overlap for the surveyed fields, and that ``\textit{bibliometrics currently offer levels of potential and methodological maturity that should induce a reconsideration and revision of their role}." Furthermore, the study shows that, with the benefit of hindsight, Italian universities, in the main, did not identify and submit for evaluation their best publications in terms of citational impact. Finally, the authors give evidence that research structures indicated as being of top quality by VTR are not necessarily also the most productive ones.

The main difference between the two mentioned previous studies and ours is the set of bibliometric indicators we have contrasted to peer judgements. Besides the journal impact factor,  we used the number of citations collected by individual papers, which directly relates to the potential impact of papers, and not to that of publishing sources, as well as the h index for relatively large publication sets. It is worth remembering that the journal impact factor was conceived as a measure of journal status, and not of impact of single papers published within it (see \citet{Ga06} and \citet{P09} for recent additions to this incessant debate). In particular, citation distributions considered in the computation of journal impact factors are always severely skewed, meaning that the majority of the papers in the journal are cited much less than the mean represented by the impact factor \citep{S92,C08}. Furthermore, we provided investigation both at the level of research disciplines (the ratings of papers) and at the level of research structures (the ratings of institutions submitting the papers). We included in the analysis also civil engineering as well as economics and statistics, for the not irrelevant fractions of submitted products that are covered by Thomson Reuters data sources. Finally, we performed different types of correlation analysis, including an intuitive probabilistic investigation.

\section{Conclusion} \label{conclusion}

We recall the research questions posed in the introduction and we propose answers based on the current investigation of the Italian research system:

\begin{enumerate}
\item \textit{Are peer review judgements and (article and journal) bibliometric indicators independent variables?}

\noindent
Both article citation and journal impact are \textit{not} independent from peer review assessment, but the correlation is positive in both cases: the higher the peer review opinion on a paper, the higher the number of citations that the paper and the publishing journal receive. Furthermore, the recently proposed h index appears to discriminate very well between sets of papers assessed with different peer judgements. It might be a viable indicator of the impact of research structures in the next editions of the evaluation exercise as soon as the average number of submitted products per structure significantly increases.

\item \textit{What is the strength of the association?}

\noindent
The correlation strength between peer assessment and bibliometric indicators is statistically significant, although not perfect. Moreover, the strength of the association varies across disciplines, and it depends also on the discipline internal coverage of the used bibliometric database (the higher the discipline coverage, the higher the reliability of citation measures). Notwithstanding, the skeptical has at disposal a few examples of papers that receive a positive peer judgement but do not collect a significant number of citations or that even sleep uncited \citep{vR04}. Furthermore, there are papers that obtain a poor judgement from peers but that rally when citations are taken into account. Even more exceptions are available when comparing peer conclusions and impact factors of journals. Nevertheless, using words of \citet{M05}, a methodology, even if provides invalid outcomes in individual cases, may be beneficial to the scholarly system as a whole.

\item \textit{Is the association between peer judgement and article citation rating significantly stronger than the association between peer judgement and journal citation rating?}

\noindent
A somewhat surprising finding of the present investigation is that the difference between the correlation strengths of article citation and journal impact factor with respect to peer assessment, although perceivable, is not as strong as one might expect.\footnote{As noticed above, the journal impact factor is a measure of journal status and not of the impact of individual papers published in the journal.}  It is worth noticing that, during the evaluation process, peer reviewers had access to the impact factors of journals that published the assessed papers, but they did not have enough information about the number of citations collected by the evaluated papers, since most of these citations were not yet mature at the time of reviewing. Therefore, peer quality opinions cannot be biased toward highly cited papers and the association between peer review and article citation is authentic.

\end{enumerate}

It is worth observing that, as already pointed out by \citet{AT04}, peer judgements and bibliometric performance measures can be expected to be  positively correlated only if the aspects assessed by the peers correspond to those reflected through bibliometric indicators. The notion of \textit{quality} assessed during peer review is perceived as a broad concept with different aspects; some of these aspects, but not necessarily all, are captured by bibliometrics. Moreover, different bibliometric measures reflect different aspects of quality, for instance, productivity, popularity, and prestige \citep{F10-JOI}.

In summary, we found a compelling body of evidence that judgements given by domain experts and bibliometric indicators are significantly positively correlated. Therefore, bibliometric indicators may be considered as \textit{approximation measures} of the inherent quality of papers, which, however, remains fully assessable only with aid of human unbiased judgement, meditation, and elaboration. We advocate the integration of peer review with bibliometric indicators, in particular those directly related to the impact of individual articles, during the next national assessment exercises. The cost effectiveness of bibliometric evaluation compared to that of peer review  would allow the evaluation of a larger sample of the universe under investigation without significant increase of costs, which is a major requirement due to the chronic national deficit and the pressing necessity of controlling public expenses in Italy.\footnote{To be sure, the cost of bibliometric evaluation is lower than that of peer review; nonetheless, every experienced bibliometrician knows that the cost to produce a reliable large-scale bibliometric assessment is far from null.} This would allow a shift from the assessment of research excellence to a more balanced evaluation of average research performance. Larger samples would, in turn, enhance the reliability of bibliometric indicators.

\bigskip
\noindent
\textbf{Acknowledgements}

\medskip
\noindent
The authors would like to thank CIVR and to its President, Prof. Franco Cuccurullo, for making available data used in this paper, through the agreement protocol between CIVR and PhD course in ``Strumenti e metodi per la valutazione della Ricerca" of the University of Chieti-Pescara.



\begin{table}
\begin{center}
\begin{tabular*}{1\textwidth}{@{\extracolsep{\fill}}lrrrr}
\textbf{MCS}  \\
\textbf{rating} &	\textbf{size} & \textbf{cites} & \textbf{IF} & \textbf{h}\\ \hline
E & 284 (36\%) & 5.52 (1.39) & 1.15 (1.02) & 16 (0.89) \\ \hline
G & 381 (48\%) & 3.31 (0.83) & 1.10 (0.98) & 13 (0.72) \\ \hline
A & 101 (13\%) & 2.61 (0.66) & 1.18 (1.05) &  7 (0.39) \\ \hline
L &  21 ( 3\%) & 2.18 (0.55) & 0.91 (0.81) &  3 (0.17) \\ \hline
&&&& \\
\textbf{PHY}  \\
\textbf{rating} &	\textbf{size} & \textbf{cites} & \textbf{IF} & \textbf{h}\\ \hline
E & 914 (52\%) & 35.30 (1.43) & 6.97 (1.20) & 85 (0.98) \\ \hline
G & 676 (38\%) & 14.19 (0.58) & 4.71 (0.81) & 40 (0.46) \\ \hline
A & 158 (9\%)  &  5.98 (0.24) & 3.10 (0.54) & 14 (0.16) \\ \hline
L &  19 (1\%)  &  5.69 (0.23) & 5.59 (0.97) &  7 (0.08) \\ \hline
&&&& \\
\textbf{CHE}  \\
\textbf{rating} &	\textbf{size} & \textbf{cites} & \textbf{IF} & \textbf{h}\\ \hline
E & 342 (32\%) & 24.72 (1.53) & 6.84 (1.32) & 42 (0.84) \\ \hline
G & 513 (47\%) & 13.54 (0.84) & 4.67 (0.91) & 34 (0.68) \\ \hline
A & 200 (18\%)  & 8.84 (0.55) & 3.57 (0.69) & 18 (0.36) \\ \hline
L &  34 ( 3\%)  & 7.57 (0.47) & 3.47 (0.67) &  9 (0.18) \\ \hline
&&&& \\
\textbf{EAS}  \\
\textbf{rating} &	\textbf{size} & \textbf{cites} & \textbf{IF} & \textbf{h}\\ \hline
E & 220 (34\%) & 10.37 (1.42) & 4.13 (1.37) & 22 (0.85) \\ \hline
G & 324 (50\%) &  6.10 (0.83) & 2.39 (0.79) & 18 (0.69) \\ \hline
A &  91 (14\%)  & 4.12 (0.56) & 2.60 (0.87) &  9 (0.35) \\ \hline
L &  16 ( 2\%)  & 6.37 (0.87) & 2.16 (0.72) &  4 (0.15) \\ \hline
&&&& \\
\textbf{BIO}  \\
\textbf{rating} &	\textbf{size} & \textbf{cites} & \textbf{IF} & \textbf{h}\\ \hline
E & 519 (33\%) & 40.86 (1.66) & 12.01 (1.42) & 75 (0.90) \\ \hline
G & 802 (51\%) & 17.97 (0.73) &  7.09 (0.84) & 49 (0.59) \\ \hline
A & 222 (14\%) & 11.59 (0.47) &  5.47 (0.64) & 21 (0.25) \\ \hline
L &  32 ( 2\%)  & 5.65 (0.23) &  5.02 (0.59) &  6 (0.07) \\ \hline
&&&& \\
\end{tabular*}
\end{center}
\caption{Peer judgement and bibliometric indicators (part I). \textbf{rating}: peer review rating (E = Excellent, G = Good, A = Acceptable, L = Limited) \textbf{size}: number of products with the given peer rating (with percentage with respect to all products),  \textbf{cites}: average number of citations of articles with the given peer rating (with ratio with respect to the average over all articles), \textbf{IF}: average impact factor of journals of articles with the given peer rating (with ratio with respect to the average over all articles), \textbf{h}: h index of articles with the given peer rating (with ratio with respect to the index over all articles).}
\label{correlation1}
\end{table}

\begin{table}[t]
\begin{center}
\begin{tabular*}{1\textwidth}{@{\extracolsep{\fill}}lrrrr}
\textbf{MED}  \\
\textbf{rating} &	\textbf{size} & \textbf{cites} & \textbf{IF} & \textbf{h}\\ \hline
E & 667  (25\%) & 47.72 (1.79) & 11.73 (1.41) & 89 (0.84) \\ \hline
G & 1314 (50\%) & 21.98 (0.82) & 7.49  (0.90) & 65 (0.61) \\ \hline
A & 492  (19\%) & 13.83 (0.52) & 6.17 (0.74)  & 36 (0.34) \\ \hline
L & 166  ( 6\%) & 14.72 (0.55) & 7.67 (0.92)  & 21 (0.20) \\ \hline
&&&& \\
\textbf{AVM}  \\
\textbf{rating} &	\textbf{size} & \textbf{cites} & \textbf{IF} & \textbf{h}\\ \hline
E & 76  (10\%)  & 16.54 (2.02) & 6.41 (2.41) & 18 (0.67) \\ \hline
G & 393 (52\%)  &  8.67 (1.06) & 2.54 (0.96) & 24 (0.89) \\ \hline
A & 218 (29\%)  &  5.15 (0.62) & 1.77 (0.66) & 15 (0.56) \\ \hline
L &  63 ( 9\%)  &  3.21 (0.39) & 1.28 (0.48) &  6 (0.22) \\ \hline
&&&& \\
\textbf{CEA}  \\
\textbf{rating} &	\textbf{size} & \textbf{cites} & \textbf{IF} & \textbf{h}\\ \hline
E & 166 (22\%)  & 5.43 (1.52) & 1.88 (1.62) & 11 (0.79) \\ \hline
G & 329 (43\%)  & 3.58 (1.00) & 1.04 (0.90) & 10 (0.71) \\ \hline
A & 217 (29\%)  & 2.29 (0.64) & 0.80 (0.70) &  7 (0.50) \\ \hline
L &  46 ( 6\%)  & 2.50 (0.70) & 0.81 (0.70) &  4 (0.29) \\ \hline
&&&& \\
\textbf{IIE}  \\
\textbf{rating} &	\textbf{size} & \textbf{cites} & \textbf{IF} & \textbf{h}\\ \hline
E & 248 (21\%) & 7.16 (1.50) & 2.03 (1.27) & 19 (0.83) \\ \hline
G & 612 (51\%) & 4.57 (0.96) & 1.56 (0.97) & 18 (0.78) \\ \hline
A & 300 (25\%) & 3.18 (0.67) & 1.33 (0.83) & 11 (0.49) \\ \hline
L &  35 ( 3\%) & 4.74 (0.99) & 1.65 (1.03) &  5 (0.22) \\ \hline
&&&& \\
\textbf{ECS}  \\
\textbf{rating} &	\textbf{size} & \textbf{cites} & \textbf{IF} & \textbf{h}\\ \hline
E & 168 (17\%) & 5.55 (1.76) &  1.31 (1.51) & 14 (0.82) \\ \hline
G & 365 (38\%) & 2.77 (0.88) &  0.75 (0.87) & 11 (0.65) \\ \hline
A & 265 (27\%) & 1.06 (0.33) &  0.58 (0.67) &  4 (0.24) \\ \hline
L & 173 (18\%) & 0.67 (0.21) &  0.48 (0.56) &  2 (0.12) \\ \hline
&&&& \\
\end{tabular*}
\end{center}
\caption{Peer judgement and bibliometric indicators (part II).}
\label{correlation2}
\end{table}

\begin{table}[t]
\begin{center}
\begin{tabular*}{1\textwidth}{@{\extracolsep{\fill}}lrrrr}
\textbf{MCS}  \\
\textbf{Rating} & \textbf{1st Q.le} & \textbf{2nd Q.le} & \textbf{3rd
Q.le} & \textbf{4th Q.le}\\ \hline
E &  27.3\% & 13.3\% & 27.3\%  & 32.0\%    \\ \hline
G & 40.0\% & 15.7\% & 24.9\%  &  19.4\%   \\ \hline
A &  48.9\% & 18.1\% & 21.3\% &  11.7\%   \\ \hline
L &   47.1\% & 35.3\% & 5.9\% &  11.8\%   \\ \hline
&&&& \\
\textbf{PHY}  \\
\textbf{Rating} & \textbf{1st Q.le} & \textbf{2nd Q.le} & \textbf{3rd
Q.le} & \textbf{4th Q.le}\\ \hline
E &  16.1\% & 20.6\% & 25.7\%  & 37.6\%    \\ \hline
G &  36.7\% & 29.2\% & 22.2\%  &  11.9\%   \\ \hline
A &  63.4\% & 25.2\% & 9.2\% &  2.3\%   \\ \hline
L &   46.2\% & 53.8\% & 0.0\% &  0.0\%   \\ \hline
&&&& \\
\textbf{CHE}  \\
\textbf{Rating} & \textbf{1st Q.le} & \textbf{2nd Q.le} & \textbf{3rd
Q.le} & \textbf{4th Q.le}\\ \hline
E &  14.3\% & 16.2\% & 25.9\%  &  43.6\%    \\ \hline
G &  35.2\% & 22.8\% & 22.8\%  &  19.8\%   \\ \hline
A &  42.4\% & 27.3\% & 23.8\% &  6.4\%   \\ \hline
L &   57.1\% & 17.9\% & 21.4\% &  3.6\%   \\ \hline
&&&& \\
\textbf{EAS}  \\
\textbf{Rating} & \textbf{1st Q.le} & \textbf{2nd Q.le} & \textbf{3rd
Q.le} & \textbf{4th Q.le}\\ \hline
E &  17.0\% & 24.0\% & 21.5\%  &  37.5\%    \\ \hline
G &  33.6\% & 27.0\% & 20.4\%  &  19.1\%   \\ \hline
A &  45.1\% & 29.6\% & 12.7\% &  12.7\%   \\ \hline
L &   37.5\% & 12.5\% & 25.0\% &  25.0\%   \\ \hline
&&&& \\
\textbf{BIO}  \\
\textbf{Rating} & \textbf{1st Q.le} & \textbf{2nd Q.le} & \textbf{3rd
Q.le} & \textbf{4th Q.le}\\ \hline
E &  11.3\% & 18.4\% & 25.3\%  &  45.0\%    \\ \hline
G &  27.7\% & 28.7\% & 27.3\%  &  16.3\%   \\ \hline
A &  50.2\% & 30.3\% & 10.9\% &  8.5\%   \\ \hline
L &   73.9\% & 13.0\% &13.0\% &  0.0\%   \\ \hline
&&&& \\
\end{tabular*}
\end{center}
\caption{Contingency table displaying the conditional distribution of article citation given peer rating (part I). Peer judgments are abbreviated as follows: E (Excellent), G (Good), A (Acceptable), L (Limited).}
\label{crossCites1}
\end{table}

\begin{table}[t]
\begin{center}
\begin{tabular*}{1\textwidth}{@{\extracolsep{\fill}}lrrrr}
\textbf{MED}  \\
\textbf{Rating} & \textbf{1st Q.le} & \textbf{2nd Q.le} & \textbf{3rd
Q.le} & \textbf{4th Q.le}\\ \hline
E & 12.3\% & 17.6\% & 25.3\%  &  44.8\%    \\ \hline
G &  25.4\% & 25.2\% & 28.3\%  &  21.1\%   \\ \hline
A &  43.4\% & 25.5\% & 19.4\% &  11.7\%   \\ \hline
L &   60.3\% & 17.8\% & 14.4\% &  7.5\%   \\ \hline
&&&& \\
\textbf{AVM}  \\
\textbf{Rating} & \textbf{1st Q.le} & \textbf{2nd Q.le} & \textbf{3rd
Q.le} & \textbf{4th Q.le}\\ \hline
E &  8.3\% & 15.3\% & 26.4\%  &  50.0\%    \\ \hline
G &  24.7\% & 25.2\% & 23.8\%  &  26.3\%   \\ \hline
A &  38.4\% & 27.0\% & 22.2\% &  12.4\%   \\ \hline
L &   52.4\% & 26.2\% & 16.7\% &  4.8\%   \\ \hline
&&&& \\
\textbf{CEA}  \\
\textbf{Rating} & \textbf{1st Q.le} & \textbf{2nd Q.le} & \textbf{3rd
Q.le} & \textbf{4th Q.le}\\ \hline
E &  18.5\% & 13.6\% & 30.9\%  &  37.0\%    \\ \hline
G &  39.9\% & 14.0\% & 23.1\%  &  23.1\%   \\ \hline
A &  54.1\% & 14.3\% & 21.4\% &  10.2\%   \\ \hline
L &   55.0\% & 15.0\% & 10.0\% &  20.0\%   \\ \hline
&&&& \\
\textbf{IID}  \\
\textbf{Rating} & \textbf{1st Q.le} & \textbf{2nd Q.le} & \textbf{3rd
Q.le} & \textbf{4th Q.le}\\ \hline
E &  25.5\% & 16.8\% & 25.0\%  &  32.7\%    \\ \hline
G &  31.9\% & 24.1\% & 21.0\%  &  23.0\%   \\ \hline
A &  44.6\% & 23.8\% &17.1\% &  14.6\%   \\ \hline
L &   52.2\% & 17.4\% & 8.7\% &  21.7\%   \\ \hline
&&&& \\
\textbf{ECS}  \\
\textbf{Rating} & \textbf{1st Q.le} & \textbf{2nd Q.le} & \textbf{3rd
Q.le} & \textbf{4th Q.le}\\ \hline
E &  14.7\% & 16.7\% & 27.3\%  &  41.3\%    \\ \hline
G &  33.7\% & 21.5\% & 23.6\%  &  21.1\%   \\ \hline
A &  54.6\% & 15.7\% & 21.3\% &  8.3\%   \\ \hline
L &   73.3\% & 6.7\% & 13.3\% &  6.7\%   \\ \hline
\end{tabular*}
\end{center}
\caption{Contingency table displaying the conditional distribution of article citation given peer rating (part II).}
\label{crossCites2}
\end{table}

\begin{table}[t]
\begin{center}
\begin{tabular*}{1\textwidth}{@{\extracolsep{\fill}}lrrrr}
\textbf{MCS}  \\
\textbf{Rating} & \textbf{1st Q.le} & \textbf{2nd Q.le} & \textbf{3rd
Q.le} & \textbf{4th Q.le}\\ \hline
E &  20.7\% & 26.2\% & 27.0\%  &  26.2\%    \\ \hline
G &  25.5\% & 26.6\% & 24.4\%  &  23.5\%   \\ \hline
A &  34.7\% & 18.9\% & 22.1\% &  24.2\%   \\ \hline
L &   29.4\% & 41.2\% & 11.8\% &  17.6\%   \\ \hline
&&&& \\
\textbf{PHY}  \\
\textbf{Rating} & \textbf{1st Q.le} & \textbf{2nd Q.le} & \textbf{3rd
Q.le} & \textbf{4th Q.le}\\ \hline
E &  15.9\% & 21.3\% & 28.4\%  & 34.3\%    \\ \hline
G &  31.7\% & 31.2\% & 25.5\%  &  11.5\%   \\ \hline
A &  54.2\% & 31.3\% & 9.2\% &  5.3\%   \\ \hline
L &   38.5\% & 38.5\% & 7.7\% &  15.4\%   \\ \hline
&&&& \\
\textbf{CHE}  \\
\textbf{Rating} & \textbf{1st Q.le} & \textbf{2nd Q.le} & \textbf{3rd
Q.le} & \textbf{4th Q.le}\\ \hline
E &  8.4\% & 16.1\% & 34.7\%  &  40.9\%    \\ \hline
G &  26.5\% & 32.9\% & 24.1\%  &  16.5\%   \\ \hline
A &  48.3\% & 31.4\% & 18.6\% &  1.7\%   \\ \hline
L &   53.6\% & 25.0\% & 17.9\% &  3.6\%   \\ \hline
&&&& \\
\textbf{EAS}  \\
\textbf{Rating} & \textbf{1st Q.le} & \textbf{2nd Q.le} & \textbf{3rd
Q.le} & \textbf{4th Q.le}\\ \hline
E &  16.5\% & 25.5\% & 22.5\%  &  35.5\%    \\ \hline
G &  28.6\% & 21.7\% & 29.6\%  &  20.1\%   \\ \hline
A &  45.2\% & 24.7\% & 17.8\% &  12.3\%   \\ \hline
L &   25.0\% & 37.5\% & 25.0\% &  12.5\%   \\ \hline
&&&& \\
\textbf{BIO}  \\
\textbf{Rating} & \textbf{1st Q.le} & \textbf{2nd Q.le} & \textbf{3rd
Q.le} & \textbf{4th Q.le}\\ \hline
E &  7.5\% & 16.8\% & 26.1\%  &  49.5\%    \\ \hline
G &  27.4\% & 31.0\% & 27.5\%  &  14.1\%   \\ \hline
A &  56.2\% & 25.9\% & 11.9\% &  6.0\%   \\ \hline
L &   60.9\% & 21.7\% & 8.7\% &  8.7\%   \\ \hline
&&&& \\
\end{tabular*}
\end{center}
\caption{Contingency table displaying the conditional distribution of journal impact factor given peer rating (part I). Peer judgments are abbreviated as follows: E (Excellent), G (Good), A (Acceptable), L (Limited).}
\label{crossIF1}
\end{table}

\begin{table}[t]
\begin{center}
\begin{tabular*}{1\textwidth}{@{\extracolsep{\fill}}lrrrr}
\textbf{MED}  \\
\textbf{Rating} & \textbf{1st Q.le} & \textbf{2nd Q.le} & \textbf{3rd
Q.le} & \textbf{4th Q.le}\\ \hline
E & 6.8\% & 17.3\% & 26.7\%  &  49.2\%    \\ \hline
G &  24.0\% & 29.0\% & 29.0\%  &  18.0\%   \\ \hline
A &  45.6\% & 29.4\% & 14.9\% &  10.2\%   \\ \hline
L &   51.0\% & 12.9\% & 12.2\% &  23.8\%   \\ \hline
&&&& \\
\textbf{AVM}  \\
\textbf{Rating} & \textbf{1st Q.le} & \textbf{2nd Q.le} & \textbf{3rd
Q.le} & \textbf{4th Q.le}\\ \hline
E &  1.4\% & 2.8\% & 19.4\%  &  76.4\%    \\ \hline
G &  15.8\% & 25.9\% & 31.6\%  &  26.7\%   \\ \hline
A &  42.0\% & 33.5\% & 17.0\% &  7.4\%   \\ \hline
L &   70.5\% & 20.4\% & 9.1\% &  0.0\%   \\ \hline
&&&& \\
\textbf{CEA}  \\
\textbf{Rating} & \textbf{1st Q.le} & \textbf{2nd Q.le} & \textbf{3rd
Q.le} & \textbf{4th Q.le}\\ \hline
E &  11.1\% & 7.4\% & 27.2\%  &  54.3\%    \\ \hline
G &  25.0\% & 24.3\% & 32.6\%  &  18.1\%   \\ \hline
A &  34.7\% & 38.8\% & 18.4\% &  8.2\%   \\ \hline
L &   38.1\% & 28.6\% & 23.8\% &  9.5\%   \\ \hline
&&&& \\
\textbf{IID}  \\
\textbf{Rating} & \textbf{1st Q.le} & \textbf{2nd Q.le} & \textbf{3rd
Q.le} & \textbf{4th Q.le}\\ \hline
E &  13.5\% & 24.0\% & 25.5\%  &  37.0\%    \\ \hline
G &  25.5\% & 23.6\% & 26.3\%  &  24.6\%   \\ \hline
A &  33.6\% & 29.5\% & 21.6\% &  15.4\%   \\ \hline
L &   39.1\% & 8.7\% & 26.1\% &  26.1\%   \\ \hline
&&&& \\
\textbf{ECS}  \\
\textbf{Rating} & \textbf{1st Q.le} & \textbf{2nd Q.le} & \textbf{3rd
Q.le} & \textbf{4th Q.le}\\ \hline
E &  5.3\% & 14.0\% & 34.7\%  &  46.0\%    \\ \hline
G &  27.2\% & 27.2\% & 27.2\%  &  18.4\%   \\ \hline
A &  45.5\% & 32.7\% & 8.2\% &  13.6\%   \\ \hline
L &   53.3\% & 33.3\% & 6.7\% &  6.7\%   \\ \hline
\end{tabular*}
\end{center}
\caption{Contingency table displaying the conditional distribution of journal impact factor given peer rating (part II).}
\label{crossIF2}
\end{table}

\begin{table}[t]
\begin{center}
\begin{tabular*}{1\textwidth}{@{\extracolsep{\fill}}lrrrrrr}
\textbf{MCS}  \\
& \textbf{cites} & \textbf{cites} & \textbf{cites} & \textbf{IF} & \textbf{IF} & \textbf{IF} \\
\textbf{ratings} &	$>$ & $<$ & $=$ &	$>$ & $<$ & $=$ \\ \hline
E $\sim$ G & 0.54 & 0.35 & 0.11 & 0.55 & 0.45 & 0.00  \\ \hline
G $\sim$ A & 0.49 & 0.36 & 0.15 & 0.53 & 0.47 & 0.00  \\ \hline
A $\sim$ L & 0.40 & 0.41 & 0.19 & 0.57 & 0.43 & 0.00  \\ \hline
&&&& \\
\textbf{PHY}  \\
& \textbf{cites} & \textbf{cites} & \textbf{cites} & \textbf{IF} & \textbf{IF} & \textbf{IF} \\
\textbf{ratings} &	$>$ & $<$ & $=$ &	$>$ & $<$ & $=$ \\ \hline
E $\sim$ G & 0.69 & 0.29 & 0.02 & 0.66 & 0.32 & 0.02  \\ \hline
G $\sim$ A & 0.66 & 0.29 & 0.05 & 0.66 & 0.33 & 0.01  \\ \hline
A $\sim$ L & 0.40 & 0.53 & 0.07 & 0.36 & 0.64 & 0.00  \\ \hline
&&&& \\
\textbf{CHE}  \\
& \textbf{cites} & \textbf{cites} & \textbf{cites} & \textbf{IF} & \textbf{IF} & \textbf{IF} \\
\textbf{ratings} &	$>$ & $<$ & $=$ &	$>$ & $<$ & $=$ \\ \hline
E $\sim$ G & 0.67 & 0.31 & 0.02 & 0.71 & 0.27 & 0.02  \\ \hline
G $\sim$ A & 0.57 & 0.39 & 0.04 & 0.68 & 0.31 & 0.01  \\ \hline
A $\sim$ L & 0.53 & 0.41 & 0.06 & 0.56 & 0.43 & 0.01  \\ \hline
&&&& \\
\textbf{EAS}  \\
& \textbf{cites} & \textbf{cites} & \textbf{cites} & \textbf{IF} & \textbf{IF} & \textbf{IF} \\
\textbf{ratings} &	$>$ & $<$ & $=$ &	$>$ & $<$ & $=$ \\ \hline
E $\sim$ G & 0.61 & 0.33 & 0.06 & 0.60 & 0.38 & 0.02  \\ \hline
G $\sim$ A & 0.56 & 0.35 & 0.09 & 0.61 & 0.38 & 0.01  \\ \hline
A $\sim$ L & 0.34 & 0.57 & 0.09 & 0.41 & 0.57 & 0.02  \\ \hline
&&&& \\
\textbf{BIO}  \\
& \textbf{cites} & \textbf{cites} & \textbf{cites} & \textbf{IF} & \textbf{IF} & \textbf{IF} \\
\textbf{ratings} &	$>$ & $<$ & $=$ &	$>$ & $<$ & $=$ \\ \hline
E $\sim$ G & 0.68 & 0.30 & 0.02 & 0.74 & 0.25 & 0.01  \\ \hline
G $\sim$ A & 0.64 & 0.33 & 0.03 & 0.68 & 0.32 & 0.00  \\ \hline
A $\sim$ L & 0.59 & 0.35 & 0.06 & 0.57 & 0.43 & 0.00  \\ \hline
&&&& \\
\end{tabular*}
\end{center}
\caption{Probability analysis of peer judgment and bibliometric indicators (part I). For each pair of adjacent peer ratings, we compute probabilities $P(c(X) > c(Y))$, $P(c(X) < c(Y))$, $P(c(X) = c(Y))$ and $P(IF(X) > IF(Y))$, $P(IF(X) < IF(Y))$, $P(IF(X) = IF(Y))$. Peer judgments are abbreviated as follows: E (Excellent), G (Good), A (Acceptable), L (Limited).}
\label{probability1}
\end{table}

\begin{table}[t]
\begin{center}
\begin{tabular*}{1\textwidth}{@{\extracolsep{\fill}}lrrrrrr}
\textbf{MED}  \\
& \textbf{cites} & \textbf{cites} & \textbf{cites} & \textbf{IF} & \textbf{IF} & \textbf{IF} \\
\textbf{ratings} &	$>$ & $<$ & $=$ &	$>$ & $<$ & $=$ \\ \hline
E $\sim$ G & 0.65 & 0.33 & 0.02 & 0.72 & 0.28 & 0.00  \\ \hline
G $\sim$ A & 0.61 & 0.36 & 0.03 & 0.65 & 0.34 & 0.01  \\ \hline
A $\sim$ L & 0.60 & 0.35 & 0.05 & 0.51 & 0.49 & 0.00  \\ \hline
&&&& \\
\textbf{AVM}  \\
& \textbf{cites} & \textbf{cites} & \textbf{cites} & \textbf{IF} & \textbf{IF} & \textbf{IF} \\
\textbf{ratings} &	$>$ & $<$ & $=$ &	$>$ & $<$ & $=$ \\ \hline
E $\sim$ G & 0.67 & 0.29 & 0.04 & 0.84 & 0.16 & 0.00  \\ \hline
G $\sim$ A & 0.58 & 0.35 & 0.07 & 0.72 & 0.28 & 0.00  \\ \hline
A $\sim$ L & 0.55 & 0.35 & 0.10 & 0.69 & 0.31 & 0.00  \\ \hline
&&&& \\
\textbf{CEA}  \\
& \textbf{cites} & \textbf{cites} & \textbf{cites} & \textbf{IF} & \textbf{IF} & \textbf{IF} \\
\textbf{ratings} &	$>$ & $<$ & $=$ &	$>$ & $<$ & $=$ \\ \hline
E $\sim$ G & 0.60 & 0.31 & 0.09 & 0.72 & 0.27 & 0.01  \\ \hline
G $\sim$ A & 0.54 & 0.31 & 0.15 & 0.62 & 0.37 & 0.01  \\ \hline
A $\sim$ L & 0.39 & 0.41 & 0.20 & 0.48 & 0.52 & 0.00  \\ \hline
&&&& \\
\textbf{IIE}  \\
& \textbf{cites} & \textbf{cites} & \textbf{cites} & \textbf{IF} & \textbf{IF} & \textbf{IF} \\
\textbf{ratings} &	$>$ & $<$ & $=$ &	$>$ & $<$ & $=$ \\ \hline
E $\sim$ G & 0.54 & 0.38 & 0.08 & 0.59 & 0.41 & 0.00  \\ \hline
G $\sim$ A & 0.53 & 0.36 & 0.11 & 0.69 & 0.41 & 0.00  \\ \hline
A $\sim$ L & 0.45 & 0.40 & 0.15 & 0.44 & 0.56 & 0.00  \\ \hline
&&&& \\

\textbf{ECS}  \\
& \textbf{cites} & \textbf{cites} & \textbf{cites} & \textbf{IF} & \textbf{IF} & \textbf{IF} \\
\textbf{ratings} &	$>$ & $<$ & $=$ &	$>$ & $<$ & $=$ \\ \hline
E $\sim$ G & 0.59 & 0.28 & 0.13 & 0.75 & 0.25 & 0.00  \\ \hline
G $\sim$ A & 0.50 & 0.25 & 0.25 & 0.63 & 0.37 & 0.00  \\ \hline
A $\sim$ L & 0.37 & 0.20 & 0.43 & 0.55 & 0.44 & 0.01  \\ \hline
&&&& \\
\end{tabular*}
\end{center}
\caption{Probability analysis of peer judgment and bibliometric indicators (part II).}
\label{probability2}
\end{table}

\end{document}